\pdfoutput=1
\documentclass[twocolumn, tighten]{aastex61}
\usepackage{amsmath}

\shorttitle{Interstellar communication. II. Application to the solar gravitational lens}
\shortauthors{Hippke}

\begin{document}
\title{Interstellar communication. II. Application to the solar gravitational lens}

\author[0000-0002-0794-6339]{Michael Hippke}
\affiliation{Sonneberg Observatory, Sternwartestr. 32, 96515 Sonneberg, Germany}
\email{hippke@ifda.eu}

\begin{abstract}
We have shown in paper I of this series \citep{2017arXiv170603795H} that interstellar communication to nearby (pc) stars is possible at data rates of bits per second per Watt between a 1\,m sized probe and a large receiving telescope (E-ELT, 39\,m), when optimizing all parameters such as frequency at 300--400\,nm. We now apply our framework of interstellar extinction and quantum state calculations for photon encoding to the solar gravitational lens (SGL), which enlarges the aperture (and thus the photon flux) of the receiving telescope by a factor of $>10^9$. 

For the first time, we show that the use of the SGL for communication purposes is possible. This was previously unclear because the Einstein ring is placed inside the solar coronal noise, and contributing factors are difficult to determine. We calculate point-spread functions, aperture sizes, heliocentric distance, and optimum communication frequency. The best wavelength for nearby ($<100$\,pc) interstellar communication is limited by current technology to the UV and optical band. To suppress coronal noise, an advanced coronograph is required, alternatively an occulter could be used which would require a second spacecraft in formation flight 78\,km from the receiver, and $\approx10$\,m in size.

Data rates scale approximately linear with the SGL telescope size and with heliocentric distance. Achievable (receiving) data rates from $\alpha$\,Cen are of order 10\,Mbits per second per Watt for a pair of meter-sized telescopes, an improvement of $10^7$ compared to using the same receiving telescope without the SGL. A 1\,m telescope in the SGL can receive data at rates comparable to a km-class ``normal'' telescope.
\end{abstract}

\section{Introduction}
One of the limiting factors in interstellar communication is the size of the receiving telescope, as discussed in paper I of this series \citep{2017arXiv170603795H}. Achievable data rates scale with the square of the diameter of a circular aperture for a wide range of parameters. One way to increase the receiver is the use of the solar gravitational lens (SGL) as a telescope. Albert Einstein predicted the bending of light around the Sun \citep{1911AnP...340..898E,Einstein1915}, which was experimentally verified only a few years later during solar eclipses \citep{1919Obs....42..119E,1920RSPTA.220..291D,1923PDAO....2..275C}. Later, \citet{1936Sci....84..506E} published a note on the focusing of starlight by the gravitational field of another star, and the improbability of observing this phenomenon by the chance alignment of two stars and the earth. The probability for galaxies is much higher, however, due to their much higher mass, and the first gravitational lens was observed by \citet{1979Natur.279..381W}.

It has been suggested to use a star as a gravitational lens for communication \citep{1979Sci...205.1133E,2011AcAau..68...76M}, and to send a sail-propelled mission to the required $>550$\,au from our Sun \citep{1994JBIS...47....3M}. Such a communication channel might be promising for interstellar probes: The photon gain is large for a given telescope aperture, and the distance is small compared to interstellar travel ($1/455$ the distance to $\alpha$\,Cen), offering precursor and/or accompanying mission options. While the benefits might be compelling, it has not been shown yet that the concept is feasible for real communication, e.g. due to the corona of the sun, which might add more noise than signal \citep{1981RaSc...16.1473H,Landis2017}. In this paper we show for the first time that SGL communication is practical. We describe the fundamental parameters of the lens for communication purpose, assess the frequency options, characterize the point-spread function (PSF) in the image plane, and discuss plausible signal and noise levels for an interstellar communication with nearby stars, which might be useful to increase data rates for a Starshot-like probe \citep{Lubin2016,2017Natur.542...20P} to $\alpha$\,Cen.

\begin{table}
\center
\caption{Baseline parameters}
\label{tab1}
\begin{tabular}{lll}
\hline
Name & Value & Explanation\\
\hline
$z$ & 600\,au & Heliocentric distance \\
$b$ & $1.05\,R_{\odot}$ & Impact parameter \\
$\lambda$ & $1\,\mu$m & Transmitter wavelength \\
$D_{\rm t}$ & 1\,m & Transmitter aperture \\
$d_{\rm SGL}$ & 1\,m & Receiver aperture \\
\end{tabular}
\end{table}

\section{Solar gravitational lens}
In this section, we will introduce the relevant parameters to use the SGL as a communication device, i.e. the lens geometry (section~\ref{sub_lens_geometry}), the possible frequency range (section~\ref{sgl_frequency_cutoff}), the light collection power (effective aperture, section~\ref{sub_light_collection_power}), and the resolution (section~\ref{sub_resolution}). We use a wavelength of $\lambda=1\,\mu$m and a heliocentric distance z= 600\,au as an example  unless noted otherwise (Table~\ref{tab1}), and discuss this choice in sections \ref{sec_noise_calc} and \ref{influence_of_wavelength}.

\begin{figure}
\includegraphics[width=\linewidth]{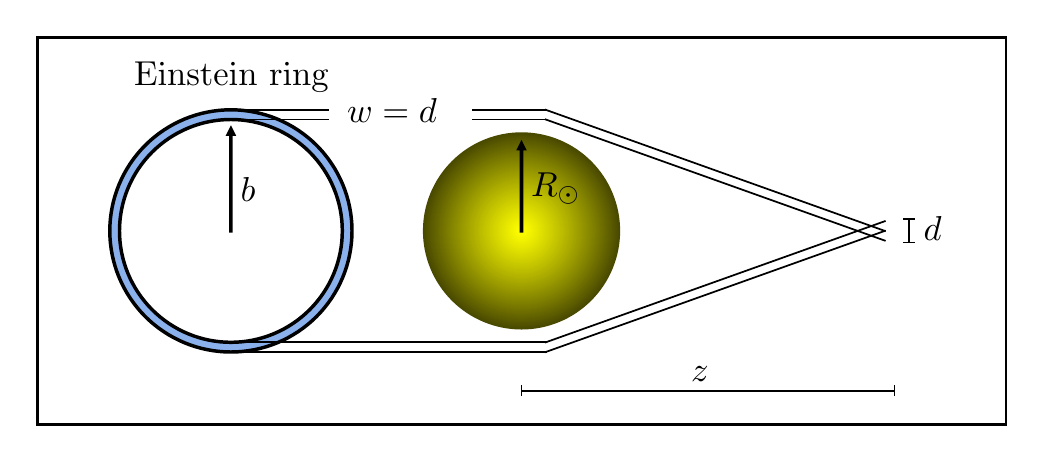}
\caption{\label{sgl_scheme}Geometry of the SGL. Rays can be traced back from the telescope diameter $d_{\rm SGL}$ (far right) at heliocentric distance $z$ to the width $w=d_{\rm SGL}$ of the Einstein ring at an angular radius $b>R_{\odot}$. Light rays follow a hyperbolic path in reality which is not shown for simplicity. The heliocentric distance $z$ is not to scale as $z_{\rm 0}=546$\,au $=1.2\times10^5R_{\odot}$.}
\end{figure}

\subsection{Lens geometry}
\label{sub_lens_geometry}
Light rays passing by a gravitational body are deflected by the body's gravitational field\footnote{In this work, we neglect Neutrinos and gravitational waves, which also pass \texttt{through} the sun, resulting in a shorter minimum focal length of $23.5\pm0.1$\,au \citep{2008ApJ...685.1297P}.}. The bending angle is inversely proportional to the impact parameter $b>R_{\odot}$ of a light ray with respect to the lens (Figure~\ref{sgl_scheme}). The inverse bending angle makes a gravitational lens different from an optical lens, which produces a single focal point. The gravitational lens, in contrast, has a focal line (a caustic) which is a focal singularity of geometric optics. A photon detector must be placed at or near this focal line. The minimum focal distance can be calculated as $z_{\rm 0}=R_{\odot}^2/2r_g \approx 546$\,au, where $r_g = 2GM_{\odot} /c^2 \approx 2,950$\,m is the Schwarzschild radius of the sun \citep{1964PhRv..133..835L,1979Sci...205.1133E}.

For larger heliocentric distances $z>z_{\rm 0}$, the impact parameter $b$ increases, i.e. the distance between the solar limb and the Einstein ring becomes larger. This relation is typically given as the minimum focusing distance as a function of the impact parameter $b$ \citep{2003MNRAS.341..577T}:
\begin{equation}
z(b)=z_{\rm 0}\frac{b^2}{R^2_\odot}
\end{equation}

which we can also solve for $b$:
\begin{equation}
b(z)=\sqrt{\frac{z}{z_{\rm 0}}}
\end{equation}

in units of $R_{\odot}$. For example, we can calculate $z(b=1.05)\approx600$\,au and $b(z=1000{\rm~au})=1.35R_{\odot}$.

The Einstein ring is very thin, $w=d_{\rm SGL}$, where $d_{\rm SGL}$ is the diameter of the circular aperture of the telescope in the focal plane (compare Figure~\ref{sgl_scheme}). A 1\,m telescope at a heliocentric distance $z=600$\,au sees the ring with an apparent width of $d_{\rm SGL}/z=4.6$\,nas, whereas the diffraction limit in the optical is $\theta=1.22\lambda/d \approx 0.25$\,arcsec, so that the ring is unresolved by a factor of $10^{-8}$; it is only resolved for $d_{\rm SGL}>10.5$\,km (at $z=600$\,au, $\lambda=1\,\mu$m) when the apparent width of the Einstein ring exceeds the angular resolution of $\theta=0.024$\,mas.

\begin{figure}
\includegraphics[width=\linewidth]{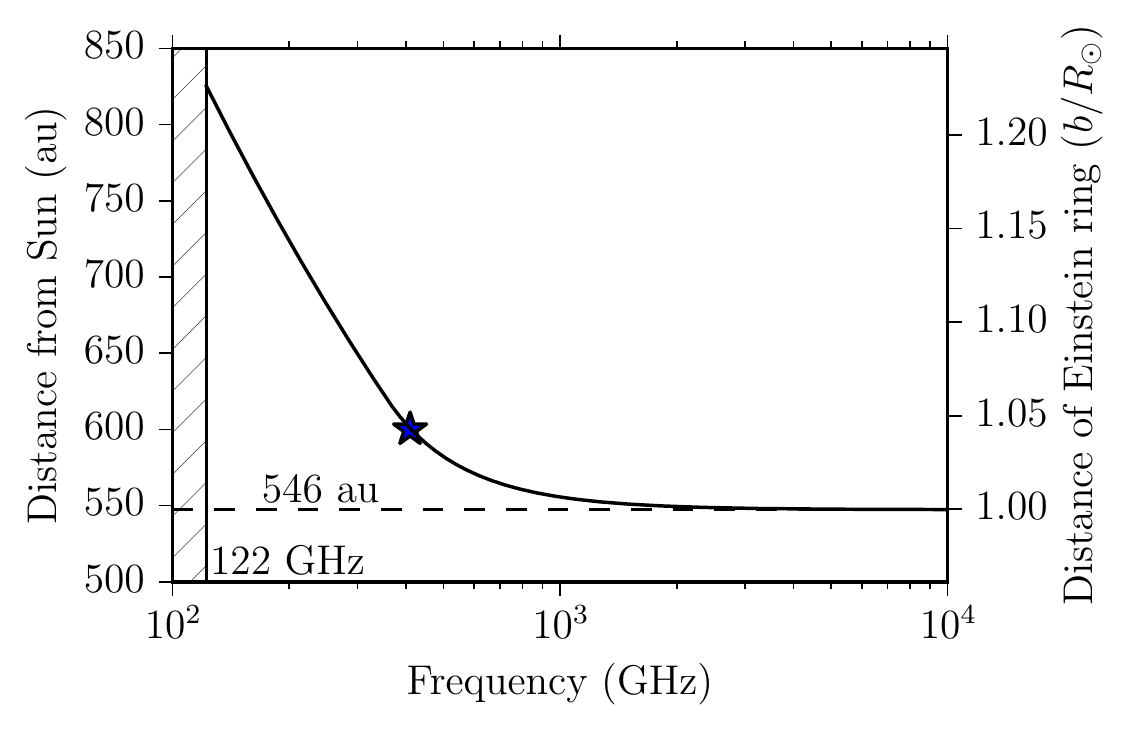}
\caption{\label{figure_crit}Critical frequency $f_{\rm crit}$ as a function of heliocentric distance $z$ (left ordinate) and impact parameter $b$ of the Einstein ring (right ordinate). The minimum frequency for which a focus occurs is $f=123.3$~GHz, lower frequencies are not focused (shaded area). For high frequencies, the closest focus approaches 546\,au (dashed horizontal line). As an example, at a distance of $z=600$~au (symbol) the Einstein ring is separated from the solar limb with $b=1.05$\,$R_{\odot}$ and the lens focuses for $f_{\rm crit}>410$\,GHz.}
\end{figure}

\subsection{Frequency cutoff}
\label{sgl_frequency_cutoff}
The solar atmosphere consists partly of a turbulent free-electron gas. The electron density at the solar photosphere is of order $10^8$\,cm$^{-3}$ \citep{1999CQGra..16.1487I}, blocking all radiation below a few hundred MHz from going through the solar atmosphere \citep{2003MNRAS.341..577T}.

A second effect is caused by refraction, which comes with a density gradient in the solar atmosphere. Even for frequencies which propagate through the gas, the anisotropy causes a refractive bending of the propagating rays, resulting in a focus shift for a given impact parameter. While gravity bends the rays in one direction, the solar plasma counters it. This has been first noted by \citet{1979Sci...205.1133E}, and first been quantified by Anderson \& Turyshev (1998, unpublished, priv. comm.).

When accounting for this refraction, the minimum focusing distance $z_{\rm min}$ can be approximated as a function of frequency $f$:
\begin{equation}
z_{\rm min} = z_{\rm 0} \,\Big(\frac{b} {R_\odot}\Big)^2
\Big[1- \frac{f^2_{0\,\rm crit}}{f^2}
\Big(\frac{R_{\odot}}{b} \Big)^{15}\Big]^{-1}~{\rm au.}
\end{equation}

A detailed analysis shows that no lensing at all occurs for frequencies $\lesssim2$\,GHz. For frequencies $f < f_{\rm crit}=123.3$\,GHz ($\lambda>2.5$\,mm, Figure~\ref{figure_crit}), the lens has no focus \citep{2003MNRAS.341..577T}. Other analyses estimate the limit at 300\,GHz \citep{2011JBIS...64...24G}, with the difference coming from different assumption for the electron density in the corona. For frequencies $f>f_{\rm crit}$, the lens quality improves and the effect of the solar atmosphere becomes negligible for $f\gtrsim1$\,THz ($\lambda\lesssim 300\,\mu$m).

\begin{figure}
\includegraphics[width=\linewidth]{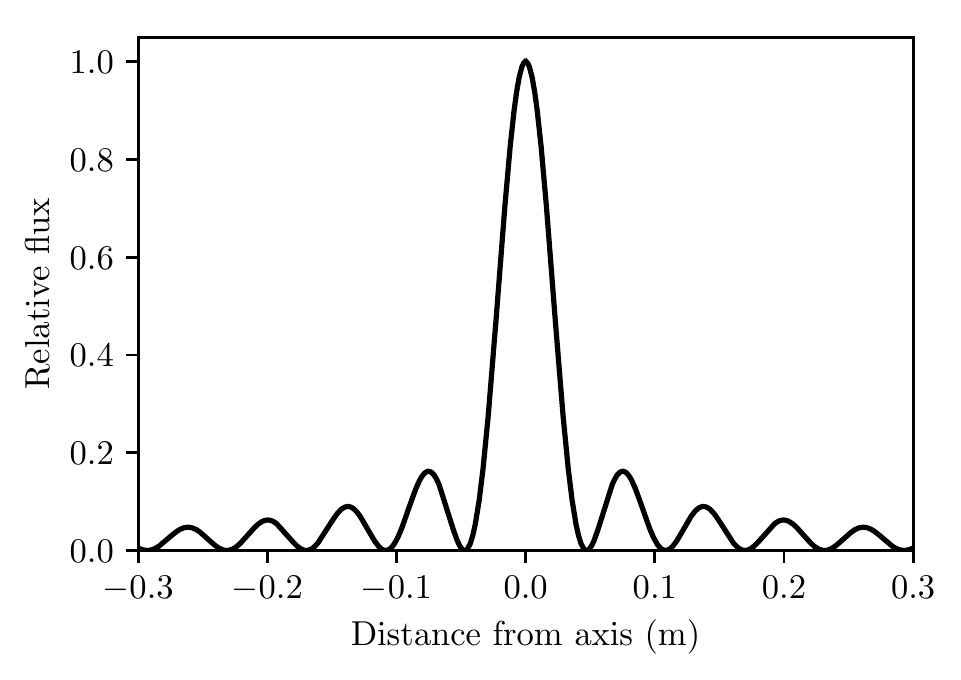}

\includegraphics[width=\linewidth]{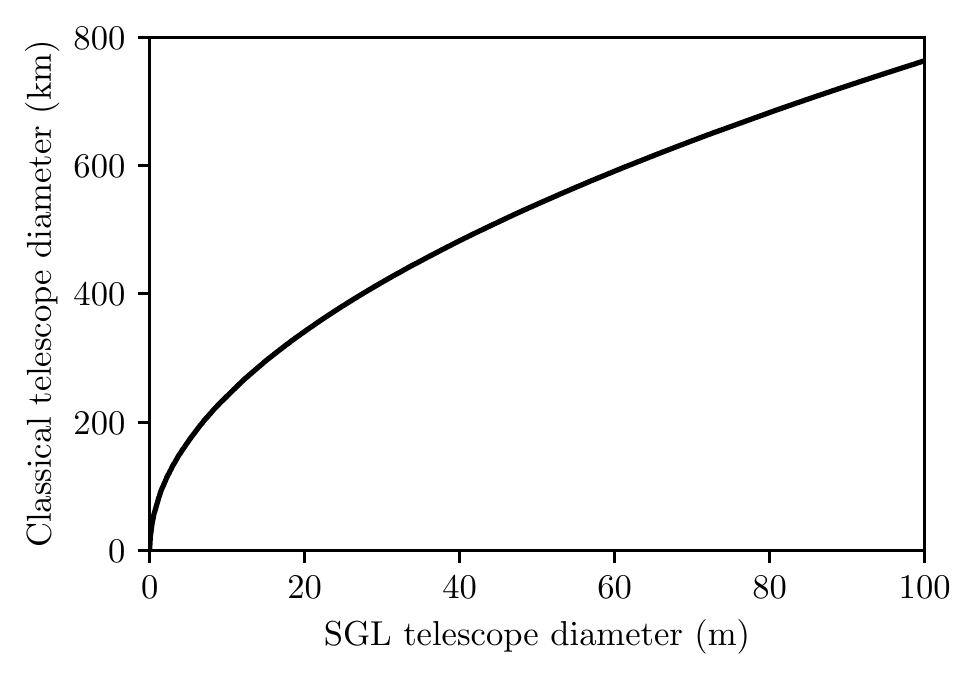}

\includegraphics[width=\linewidth]{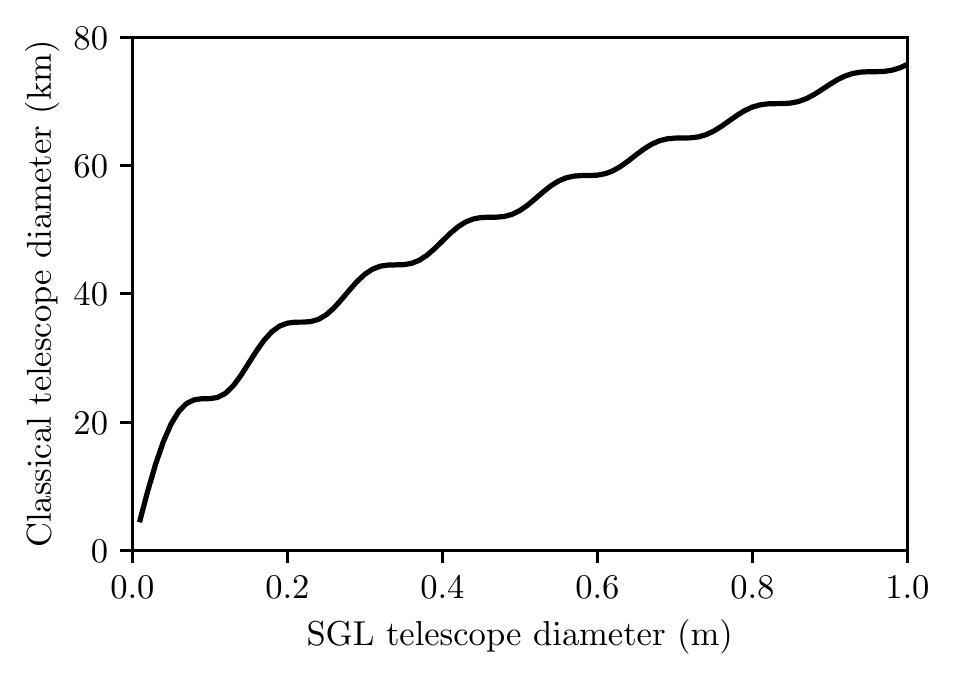}
\caption{\label{figure_bessel}Top: Relative PSF flux versus distance from the axis for $\lambda=1$\,$\mu$m, $z=600$\,au. Middle: Integration over the PSF to compare the corresponding size of a classical telescope with one in the SGL. Bottom: Zoom for $0<d_{\rm SGL}<1$\,m, showing the influence of the PSF on the increase in gain.}
\end{figure}

\subsection{Light collection power}
\label{sub_light_collection_power}
The area covered by the (very thin) Einstein ring is the effective aperture of the SGL telescope. The area of this ring can be calculated as $A_{\rm ER}=\pi ((b + w) ^2 - b^2)$ where $b$ is the inner edge of the ring, and $w$ its width. The width of the Einstein ring is so small compared to the radius that this can without any loss of accuracy be more simply expressed as $A_{\rm ER}=\pi bw$. The flux in the image plane is not uniform; it follows an Airy pattern, despite the caustic nature of gravitational lensing \citep[e.g., ][their Eq. 7-22]{schneider2013gravitational}. The distance $\rho$ from the center of the peak to the first null of the PSF can be approximated as \citep[][their Eq. 142]{2017PhRvD..96b4008T}

\begin{equation}
\rho\simeq 4.5~\Big(\frac{\lambda}{1\,\mu{\rm m}}\Big)\frac{b}{R_\odot}~{\rm cm}.
\end{equation}

For example, for $\lambda=1$\,$\mu$m and $b/R_{\odot}=1.05$, the distance from the maximum to the first null is 4.7\,cm (Figure~\ref{figure_bessel}, top). For point sources, we can calculate the flux density in the image plane as a function of wavelength $\lambda$, heliocentric distance $z$ and distance from the axis $\rho$ following \citet[][their Eq. 135]{2017PhRvD..96b4008T}:

\begin{eqnarray}
\mu=\frac{4\pi^2}{1-e^{-4\pi^2 r_g/\lambda}}\frac{r_g}{\lambda}\, J^2_0\Big(2\pi\frac{\rho}{\lambda}\sqrt{\frac{2r_g}{z}}\Big)
\end{eqnarray}

where $J_0$ is a Bessel function of order zero. The SGL PSF falls off more slowly compared to a traditional (optical lens) PSF, which is proportional to $J^2_1(2\sqrt{x})/x^2$. The magnification also depends (weakly) on the heliocentric distance $z$ as $\theta_0 \sqrt{{z}_0/{z}}$ \citep{2017PhRvD..96b4008T}.

The argument of the Bessel function can be expressed in the more traditional form of a PSF. With the quantities involved, we obtain \citep[][their Eq. 138]{2017PhRvD..96b4008T}

\begin{equation}
2\sqrt{x}=53.34\Big(\frac{1\,\mu{\rm m}}{\lambda}\Big)\Big(\frac{\rho}{1~{\rm m}}\Big)\frac{R_\odot}{b}.
\end{equation}

The SGL PSF gain has a maximum on the axis \citep{1975Ap&SS..34L...7B}
\begin{eqnarray}
\mu_{max}=4\pi\frac{r_g}{\lambda}.
\end{eqnarray}

The collected flux of a telescope in the SGL with diameter $d_{\rm SGL}$ can be calculated by spherically integrating over the PSF. For a telescope with its center on the axis, the average gain for point sources is \citep[][their Eq. 143]{2017PhRvD..96b4008T}

\begin{eqnarray}
\begin{aligned}
\label{eq_mu}
\bar{\mu}(z,d_{\rm SGL},\lambda)=\frac{4\pi^2}{1-e^{-4\pi^2 r_g/\lambda}}\frac{r_g}{\lambda} \times \\ \Big\{ J^2_0\Big(\pi\frac{d_{\rm SGL}}{\lambda}\sqrt{\frac{2r_g}{z}}\Big)+J^2_1\Big(\pi\frac{d_{\rm SGL}}{\lambda}\sqrt{\frac{2r_g}{z}}\Big)\Big\}.
\end{aligned}
\end{eqnarray}

The flux on the aperture in the SGL can be compared to the one for classical telescope outside of the SGL. For example, for $d_{\rm SGL}=1$\,m, $\lambda=1$\,$\mu$m, $z=600$\,au, we have $\bar{\mu}=2.87\times10^9$.

We can calculate the corresponding circular aperture size $D_{\rm classical}$ of a telescope outside of the SGL compared to a telescope of size $d_{\rm SGL}$ in the SGL by multiplying the averaged gain over the SGL aperture and solving for $D_{\rm classical}$:
\begin{equation}
\label{aperture_size}
D_{\rm classical} = 2 \sqrt{\frac{\bar{\mu} d_{{\rm SGL}}^2}{4}} = d_{\rm SGL} \sqrt{\bar{\mu}}.
\end{equation}

\begin{figure}
\includegraphics[width=\linewidth]{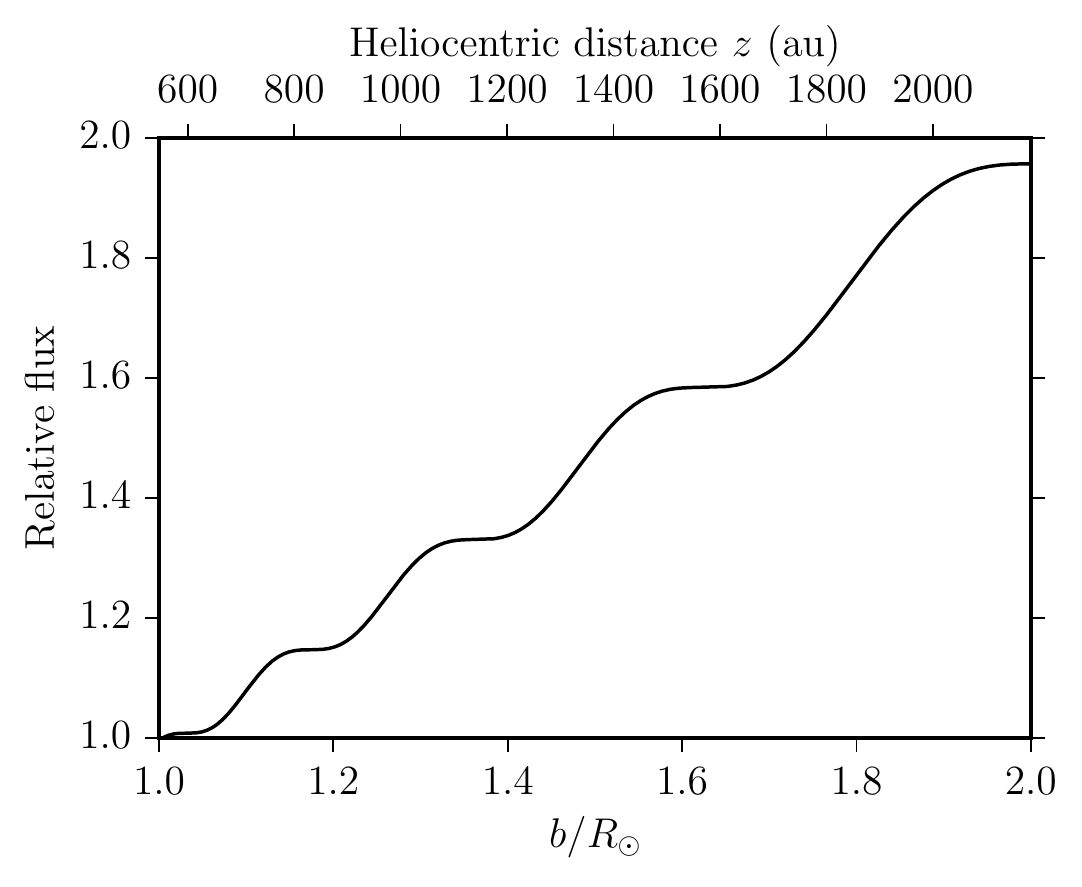}
\caption{\label{figure_z}Influence of heliocentric distance $z$ on the normalized flux collected by a telescope with a constant aperture size $d$ in the focal plane. Increasing the distance from $z_0=546$\,au to $z=2,200$ au increases the number of photons collected by a factor of $\approx$2.}
\end{figure}

The $d_{\rm SGL}=1$\,m telescope at $z=600$\,au collects as many photons as a $D_{\rm classical}=75.75$\,km (kilometer) telescope (Figure~\ref{figure_bessel}, middle, bottom). The influence of $z$ is within a factor of $\approx2$ between $z_0=546$\,au and $z=2,200$\,au (Figure~\ref{figure_z}). These calculations using the point spread function to estimate photon count rates are valid for the case where the projected size of the transmitter in the image plane is smaller than the point spread function, as will be discussed in section~\ref{sub_resolution}.

\subsection{Resolution}
\label{sub_resolution}
The resolution of the SGL is so high that nearby exoplanets (and even large telescope structures located at or around such planets) are spatially resolved. For resolved sources, the image produced in the focal plane of an SGL can be calculated from geometrical optics, where the image is smaller than the object by $R/z$, with $R$ as the distance to the object. For $z=600$\,au and our closest neighbor $\alpha$\,Cen~C, $R=1.3$\,pc \citep{2016A&A...594A.107K,2017A&A...598L...7K}, we get $R/z=455$. Thus, an earth-mass exoplanet around Proxima Centauri \citep{2016Natur.536..437A} with an exemplary earth-like diameter of 12,756\,km would appear with a size of 28.5\,km in the image plane; a 1\,m telescope in the image plane would see an area of 455\,m on the planet. The high magnification causes problems for imaging \citep[as discussed in][]{Landis2017}, because the observations must resemble a raster-scan while the planet itself rotates (around its star, and around its axis). For a 1\,m telescope, the 455\,m area of an exoplanet with an orbit of 1\,au traverses the field-of-view in 33\,ms.

A project like ``starshot'' might send many probes within a short time frame, so that a fleet of transmitters could be formed near Proxima. The maximum sky-projected distance between the probes needs to be $<d_{\rm SGL}R/z$ (455\,m for a 1\,m SGL telescope) to have all individual PSF peaks in the field of view. From probes with larger separations, only sidelobes can be received, which contribute only marginally to the total signal. The exact placement of probes inside this 455\,m disk is irrelevant, because the PSF is positive for all distance $\rho$ from the axis, so that the summed (received) PSF is additive.

If the transmitter in this example (e.g. located on an exoplanet around $\alpha$\,Cen) has $D_{\rm t}\geq445$\,m, it appears as an extended source in the image plane of the receiver in the SGL. Then, the brightness distribution places 50\% of the flux into the inner 50\% of the source's diameter, although the inner half contributes only 25\% of the area \citep{Landis2017}.

Consequently, for large (km-size) nearby (pc) senders, the size of the receiver in the SGL can be optimized depending on $\lambda$, $z$, and $D_{\rm t}$. For optical to near-infrared wavelengths, the optimal receiver size will scale as a few times $z/R$, so that a km-scale transmitter and a meter-sized receiver in the SGLs are a good combination for the nearest stars.

\section{Signal}
We will now count the number of photons from a distant source (``signal''), as received by a telescope in the SGL. Together with the noise photons (section~\ref{sec:noise}) and the capacity (in bits per photon, section~\ref{sec_capacity}) we can then estimate the maximum data rate within Holevo's bound \citep{holevo1973bounds}.

\begin{figure}
\includegraphics[width=\linewidth]{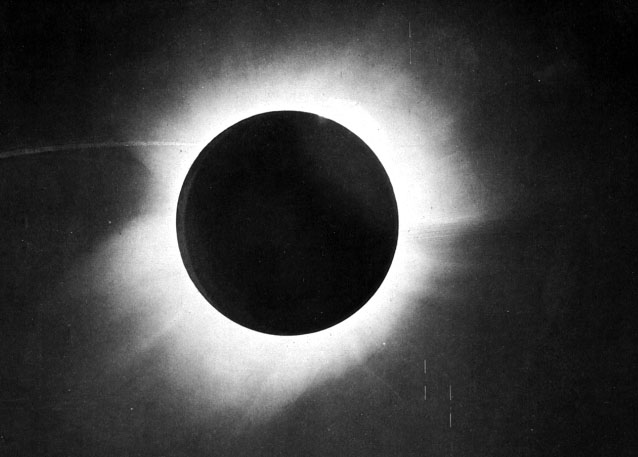}
\caption{\label{corona.jpg}Solar eclipse photograph from \citet{1920RSPTA.220..291D} which was used to verify ``Einstein's new law of gravitation'' by measuring the displacement of stars near the limb. The corona is clearly visible, but the faint stars in it require a loupe to be seen.}
\end{figure}

\subsection{Photon flux}
In paper I \citep{2017arXiv170603795H}, we have defined the photon flux $P_{\rm r}$ received by a telescope with aperture $D_{\rm r}$ from a distance\footnote{In this paper we denote distance as $R$ instead of $d$, in contrast to \citet{2017arXiv170603795H}, to avoid confusion with the aperture size.} $R$ with power $P_{\rm t}$:
\begin{equation}
\label{photon_flux}
P_{\rm r}= \frac{P_{\rm t} D_{\rm t}^2 D_{\rm r}^2}{4 h f (Q_{\rm R} \lambda)^2 R^2} ({\rm s}^{-1})
\end{equation}

where $Q_{\rm R}\approx1.22$ for a diffraction limited circular transmitting telescope of diameter $D_{\rm t}$ \citep{rayleigh}, $h$ as Planck's constant ($\approx6.626\times10^{-34}$\,J\,s), and wavelength $\lambda=c/f$ with $c$ as the speed of light in vacuum ($299,792,458$\,m\,s$^{-1}$).

We can insert the aperture size of our SGL telescope from Eq.~\ref{aperture_size}, $D_{\rm r}=d_{{\rm SGL}} \sqrt{\bar{\mu}} \Leftrightarrow D_{\rm r}^2 = d_{{\rm SGL}}^2 \bar{\mu}$ into Eq.~\ref{photon_flux}:

\begin{equation}
\label{photon_flux_2}
P_{\rm r}= \frac{P_{\rm t} D_{\rm t}^2 d_{\rm SGL}^2 \bar{\mu}}{4 h f (Q_{\rm R} \lambda)^2 R^2} ({\rm s}^{-1})
\end{equation}

and calculate the number of photons received by the telescope in the SGL, where $\bar{\mu}$ from Eq.~\ref{eq_mu} accounts for the SGL gain and the PSF.

For example, a pair of classical 1\,m telescopes at a distance of 1.3\,pc transmitting at $\lambda=1$\,$\mu$m can exchange $5\times10^{-4}$ photons per Watt per second (neglecting losses other than free-space). The same 1\,m receiving telescope in the SGL with its equivalent aperture of $D_{\rm classical}=d_{\rm SGL} \sqrt{\bar{\mu}}=75.75$\,km collects $3\times10^6$ photons per second per Watt, a gain of $\bar{\mu}=5.74\times10^9$ (section~\ref{sub_light_collection_power}).

\section{Noise}
\label{sec:noise}

\begin{figure*}
\includegraphics[width=.5\linewidth]{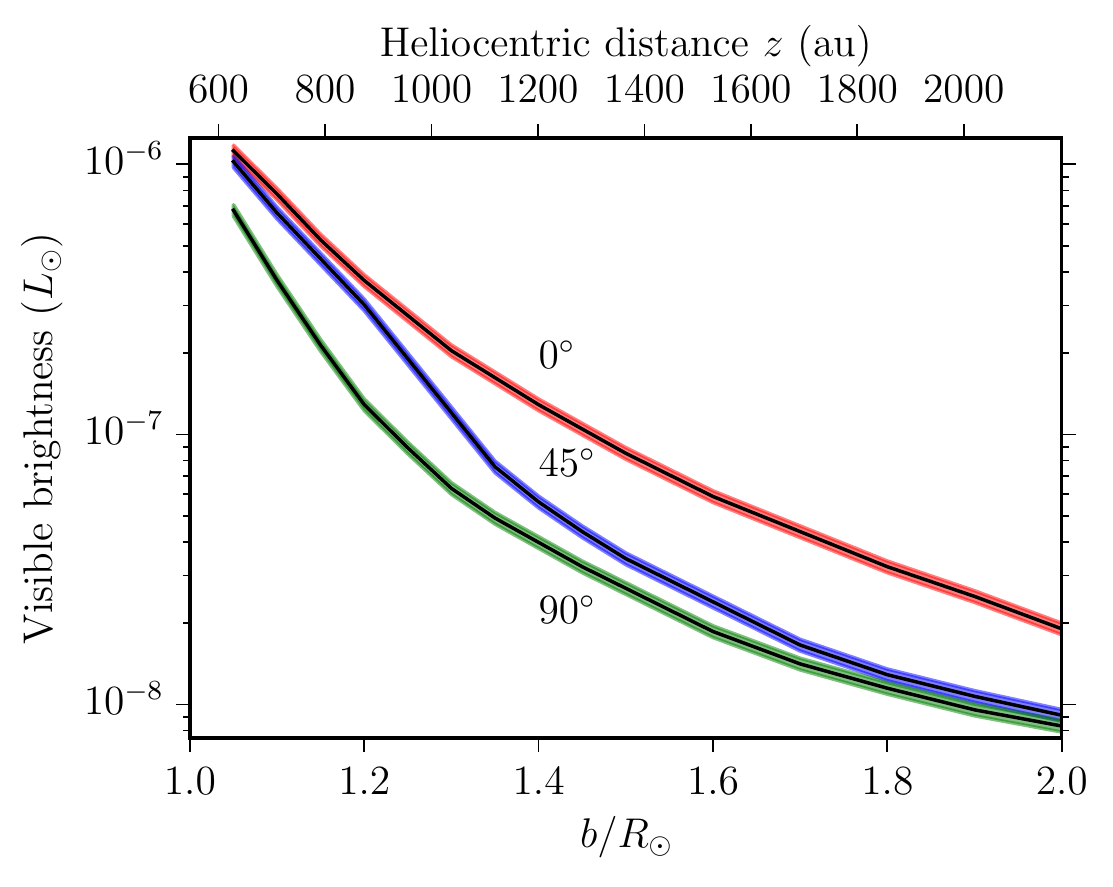}
\includegraphics[width=.5\linewidth]{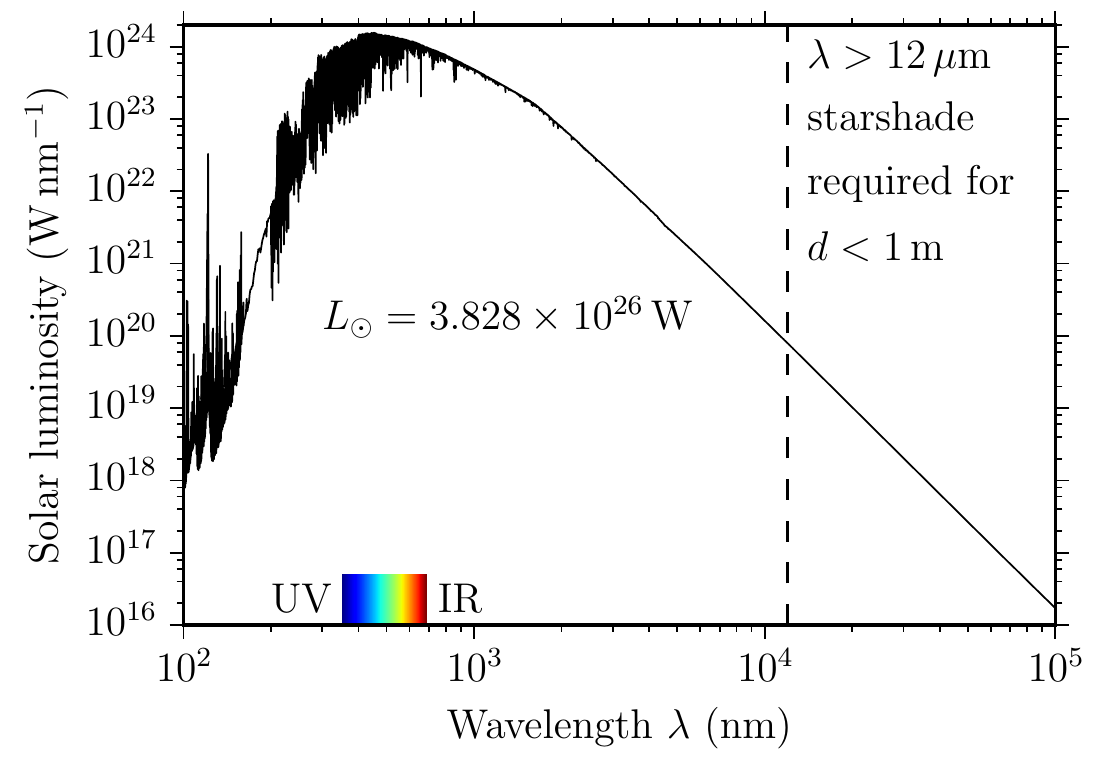}
\caption{\label{figure_corona}Left: The visible brightness of the solar corona decreases with radial distance, depending on heliographic latitude. Data from \citet{1961ZA.....53...81W}. Right: The spectrum of the solar disk peaks at visible wavelengths, whereas the coronal light is flat between $0.3<\lambda<1$\,$\mu$m (not shown). See text for discussion.}
\end{figure*}

\subsection{The solar corona}
\label{sub_corona}
After Albert Einstein's prediction that ``light bends'' \citep{1911AnP...340..898E,Einstein1915}, two expeditions traveled to Africa and Brazil to observe the next solar eclipse in 1919. The displacement of the stars near the solar limb by up to 1.75\,arcsec was clearly detected and ``Einstein's new law of gravitation'' verified \citep{1920RSPTA.220..291D,1923PDAO....2..275C}. A major obstacle in the measurement of star locations in these photographs was the solar corona, whose brightness washed out stars dimmer than 10th magnitude (Figure~\ref{corona.jpg}).

The spectrum of the solar corona is different from that of the sun. It consists of three components: the K-corona created by sunlight scattering of free electrons, in which absorption lines are completely smeared due to Doppler broadening. The F-corona is created by light reflected off dust particles, and it contains the Fraunhofer absorption lines. The E-corona originates from spectral emission produced by ions in the coronal plasma \citep{1964P&SS...12...55S}. Overall, the coronal spectrum is essentially flat from $0.3-1$\,$\mu$m within a factor of $\approx2$ and decreases by an order of magnitude at 3\,$\mu$m \citep{1998EP&S...50..493K}.

The spectrum has a number of prominent emission lines with flux levels $\approx4\times$ the continuum \citep{1996ApJ...456L..67K}, e.g. the Fe~XIII line at 1.0747\,$\mu$m; lines which should be avoided for communication as they are strong noise sources. There are also possible, but not well characterized absorption lines, e.g. near $10.8\,\mu$m, attributed to the minimum emissivity wavelength of silicate-type materials \citep{1974A&A....37...81L}. A high resolution spectrum of the solar corona for $1.05<b<2$ would be very helpful for a detailed analysis. We did not find a publication on such a spectrum despite an in-depth literature research.

The brightness of the corona can be approximated as $L=b^{-6}\times10^{-6}L_{\odot}$ for $1.05<b<2$ ($600<z<2200)$ for wavelengths $0.3<\lambda<3$\,$\mu$m and low heliographic latitudes to within 10\%, which is sufficient for our purposes \citep{1998A&AS..127....1L,1998EP&S...50..493K}. The brightness in visible light is shown in Figure~\ref{figure_corona} (left) for different radial distances, with the corresponding heliocentric distance $z$ for each impact parameter shown on the upper abscissa. The brightness shown in this Figure is in between the one for the solar minimum \citep[integrated $0.52\pm0.08 \times 10^{-6}L_{\odot}$, ][]{1977MitSZ.356.....D} and maximum \citep[$1.29\pm0.05 \times 10^{-6}L_{\odot}$, ][]{1982A&A...112..241D}.

In contrast to classical interstellar communication, where stellar light is the dominant noise source \citep[][Figure~\ref{figure_corona}, right]{2017arXiv170603795H}, we can not use low-flux (low-noise) Fraunhofer absorption lines for communication, because these are smeared.

The coronal spectrum at wavelengths shorter than $300$\,nm (UV, X-Ray, $\gamma$-Ray) will be neglected here for communication purposes because of the inability of the current technology to efficiently focus these wavelengths \citep{2017arXiv170603795H}. While the SGL would be the perfect telescope to focus such a communication for the receiver, we do not currently possess the technology to focus it for transmission, particularly not on-board a small, lightweight probe. This aspect will be covered in detail in paper III of this series, when we relax technological constraints.

\subsection{Observation strategy}
\label{observation_strategy}
Our telescope in the SGL has to monitor the Einstein ring around the sun, which delivers the signal. Consequently, we have to point our telescope directly at the sun. Doing this without a filter from a heliocentric distance of $z=1$\,au would result in a loss of the manufacturer's warranty. Further out in the SGL, however, the solar flux $F$ scales with the inverse square law, so that $F=L_{\odot} = 1,361$\,Wm$^{-2}$ at 1\,au is reduced to $F_{\rm dist}=F/4\pi z^2 = 3.8$\,mW\,m$^{-2}$ at $z=600$\,au. While this should not hurt the telescope, it is still an enormous source of noise, amounting to a flux of $\approx9.6\times10^{16}$\,$\gamma$\,s$^{-1}$\,m$^{-2}$, and would prohibit all communication if left untreated. To reduce noise, the light from the disk can be blocked either with a coronograph or an occulter.

\begin{figure*}
\includegraphics[width=\linewidth]{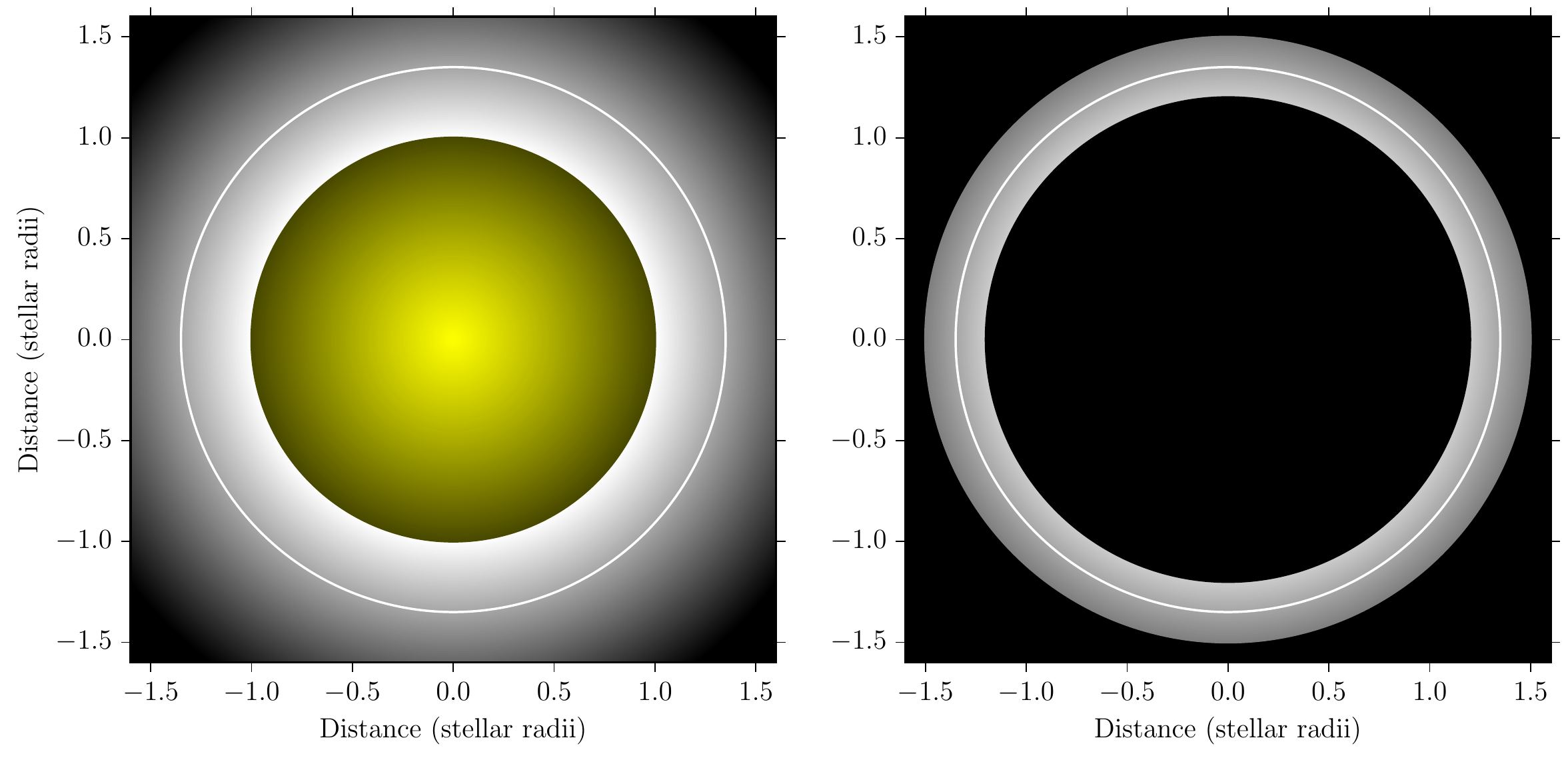}
\caption{\label{sun_double}Left: View of the sun for a probe at $z=1000$\,au. The sun has an apparent width of 2.1\,arcsec, and the Einstein ring is located at $b/R_{\odot}=1.35$, surrounded by the corona. Right: Situation with an annular aperture covering the disk and most of the corona (the noise), leaving a ring-shaped area $A_{\circ}$ centered on the Einstein ring (the signal). The width of this ring-shaped area is diffraction limited with $\theta=0.2$ arcsec for $d_{\rm SGL}=1$\,m at $\lambda=1$\,$\mu$m.}
\end{figure*}

\subsection{Coronograph}
\label{sec:coronograph}
A coronograph can block most light \citep[$10^{-9}$,][]{2006ApJS..167...81G,2015RAA....15..453L} from the solar disk. Ideally, the coronograph would have an annular aperture, so that only a ring-shaped region around the sun remains visible (Figure~\ref{sun_double}). Such a device is not known yet. For cases where the signal flux is within an order of magnitude of the coronal noise (section~\ref{sgl_aperture}), so that it is not totally overpowered, it would be fully sufficient to only block the solar light using a classical (inner working angle) coronograph. In our standard case, this regime begins at $P_{\rm t}>0.1\,$W for narrow (nm) bandwidths.

If an annular coronograph is available, its aperture must be centered on the Einstein ring. The width of the slit is at minimum defined by diffraction. Reducing the flux from the disk by $10^{-9}$ is sufficient, because at this point the noise is dominated by the light from the solar corona, about $10^{-6}$ of the brightness of the solar disk. In practice, the inner borders of the ring-shaped area have no sharp transition, but one that is determined by the (telescope's) PSF and the coronograph implementation. We neglect this difference here and assume a sharp transition, resulting in a conservative (upper limit) estimate of resulting noise.

It is also worth noting that a coronograph reduces the optical throughput of a system.

As the Einstein ring is close to the limb, the resolution of the telescope needs to be at the same level in order to separate the disk from the Einstein ring. The diffraction-limited resolution equals the width of the solar disk for $d=1$\,m and $\lambda=12$\,$\mu$m, rendering a coronograph insufficient even at large $z$ for longer wavelengths at this mirror size (Figure~\ref{figure_corona}, right). Then, the noise from the solar disk itself also enters as noise into the aperture. The solar flux at $12$\,$\mu$m is $10^{20}$\,W\,nm$^{-1}$, which is over 4 orders of magnitude less than at the peak near 400\,nm. However, the coronograph can block the flux from the disk when observing at 400\,nm, and the flux from the corona is $>10^6\times$ less, so that the received noise at $12$\,$\mu$m is actually 2 orders of magnitude higher than at 400\,nm. Given the flat coronal spectrum, the situation is even worse, amounting to a total of 2--3 orders of magnitude more noise at $12$\,$\mu$m. Going into the THz and microwave regime reduces the noise flux, but allows for less efficient focusing on the transmitter side, as will be discussed in section~\ref{sec_noise_calc}.

\subsection{Occulter}
\label{sec:occulter}
The idea to block starlight with an occulter (a starshade) was first proposed by \citet{10.2307/27838457}, and a number of studies have been conducted in the last decade to analyze its performance for direct imaging of exoplanets. Its main strength is that the contrast and inner working angle are decoupled from the telescope aperture size. The major disadvantage is observational, as the shade needs to be slewed to every new target, which takes time and fuel. For our case, where we send probes on a straight path, an occulter could make sense by sending a second probe shortly after the first, on the same trajectory, casting a shade on the first, the telescope.

The problem of minimizing diffraction from the edge of the shade has led to the development of a series of optimized shapes, converging on an ``offset hypergaussian'' which uses petals to approximate a circularly symmetric function. Visually, such an occulter resembles a flower-shaped disk. Because of the complexity of finding an optimum shape for our specific use case, including wavelength, the design of the best occulter is outside of the scope of this paper. For simplicity, we will discuss an order-of-magnitude estimate using established starshade models, following \citet{2007ApJ...665..794V,2009arXiv0912.2938C} and \citet{2011ApJ...738...76C}. We assume, as discussed in the previous section, that the transmitter power is sufficiently high so that occulting most of the corona is not required.

For our example, we keep the telescope diameter as $d_{\rm SGL}=1$\,m at $z=600$\,au. From this heliocentric distance, the sun subtends to 3.19\,arcsec, and the Einstein ring is at $b=1.05$, or $\alpha=1.67$\,arcsec from the axis. This is where we need the inner working angle of the starshade. If we restrict the size of the occulter to something reasonable, e.g. $d_{\rm occult}=10$\,m, and place the slit at 5\,m (half the occulter's width), the distance between the telescope and the occulter must be $z_{\rm dist}=d_{\rm occult}/ \alpha \approx 78$\,km. If both spacecrafts move further out, this distance needs to change, but we neglect the issue here. Using an optimized geometry, a working angle of 0.1\,arcsec with contrast of $10^{-10}$ or better is achievable, equal to or better than the coronograph.

\subsection{Other noise sources}
If the transmitter is perfectly aligned with a star along our line of sight, the stellar noise photons will be blended into the transmitter signal, reducing data rate to extremely low levels \citep{2017arXiv170603795H}. Fortunately, as explained in section~\ref{sub_resolution}, the magnification (resolution) of the lens in the image plane is very high, e.g. on the km-scale for nearby (pc) stars. We can estimate the noise from the star for a sky-projected separation by calculating the gain as a function of distance from the axis, $\mu(\rho)$, for the PSF as shown in Figure~\ref{figure_bessel}. A contrast ratio of $10^{-9}$ for $\lambda=1$\,$\mu$m and $z=600$\,au occurs at $\rho=6,200$\,km from the axis. Using $R/z=455$, we can estimate the necessary separation for the transmitter at a distance of 1.3\,pc as $455\times6,200=2.8\times10^9$\,m or $4 R_{\odot}$. Hence, if the probe keeps further away from the target star than a few solar radii, stellar noise will negligible.

Stray light on the direct path (e.g., from stars and planets) can be reduced using standard telescope design principles, such as tubes or baffles.

\subsection{Noise calculations}
\label{sec_noise_calc}
We will now calculate the integrated flux over the ring-shaped area which covers the Einstein ring. This area is a function of heliocentric distance $z$, telescope diameter $d_{\rm SGL}$ and wavelength $\lambda$. It is important to understand that the impact parameter $b$ moves further out with larger distances $z$ (which decreases noise following a power-law), but for a given diffraction-limited telescope aperture $d_{\rm SGL}$ (and constant wavelength), the area increases (which increases noise). Focusing improves for shorter wavelengths, decreasing the area and thus noise.

The apparent width of the sun, as a function of heliocentric distance $z$, is $w_{\odot}=3600\times180/\pi \times 2 R_{\odot} / z$\,arcsec, and its apparent area is

\begin{equation}
A_{\odot}=\pi (w_{\odot}/2)^2=\pi w_{\odot}^2/4~{\rm (arcsec^2)}.
\end{equation}

The resolution of our telescope, and therefore the width of the ring-shaped area, is $w_{\circ}=Q_{\rm R} \lambda D$. As for the Einstein ring ($A_{\rm ER}$), we have the overlay ring as $A_{\circ}=\pi ((r + w_{\circ}) ^2 - r^2)$, where $r=bw_{\odot}/2$ shall be the distance from the solar disk's center, in arcsec. The ratio of $A_{\circ}/A_{\odot}$ multiplied with the average brightness is the total flux of this area:

\begin{equation}
L_{\circ}= \frac{\pi ((r + w_{\circ}) ^2 - r^2)}{\pi w_{\odot}^2/4} \times b^{-6}\times10^{-6}L_{\odot}
\end{equation}

For example, at $z=600$\,au, the sun subtends to a diameter of 3.19\,arcsec, so that its area is $A_{\odot}=8.03$\,arcsec$^2$. A $d_{\rm SGL}=1$\,m telescope observing at $\lambda=1$\,$\mu$m has a resolution of $\theta=0.252$\,arcsec. The Einstein ring is at a distance of $b=1.05$ (or $r=1.67$\,arcsec from the solar center). Then, we have the ring area as $A_{\circ}=2.84$\,arcsec$^2$, or $A_{\circ}/A_{\odot}=0.354$ of the solar disc. The average brightness of the ring is $L=1.05^{-6}\times10^{-6}=7.59\times10^{-7}L_{\odot}$; integrated over the area of the ring the total flux is $2.66\times10^{-7}L_{\odot}$. This is the amount of noise which is visible to the telescope from the solar corona.

The noise flux $F$ of this ring-shaped area scales to distance $z$ with the inverse square law, so that $F_{\rm dist}=F/4\pi z^2 = 2.66\times10^{-36}$\,$L_{\odot}$. Taking the solar luminosity as $L_{\odot}=3.9\times10^{26}$\,W, we have $1.02\times10^{-9}$\,W (or Js in SI units).

This is the total noise, but our telescope can use a narrow filter. Assuming a (simplified) flat coronal spectrum for wavelengths $0.3<\lambda<3$\,$\mu$m (and negligible flux outside of this band), a filter bandwidth of $1$\,nm lets pass $1/2700$ of the total flux, or $3.79\times10^{-13}$\,Js.

For a fly-by mission without deceleration, the change in velocity due to the gravitational influence of the target star causes a Doppler shift in the transmitter wavelength as seen by the receiver. If the receiver uses a constant filter, the Doppler shift must be smaller than the filter bandwidth. We simulated the velocity change of a fly-by mission using the numerical integrator from \citet{2017ApJ...835L..32H,2017AJ....154..115H}. For a close ($10\,R_{\odot}$) fly-by at the most massive star (A) of the stellar triple Alpha Cen at a speed of 20\%\,c, the maximum velocity change is 300\,m\,s$^{-1}$ and decreases to 5\,m\,s$^{-1}$ in the case of a 1\,au closest stellar encounter. Such a velocity change results in Doppler shifts $\ll 1$\,nm.

We can calculate the number of photons in this band by taking $E=hf$, so that $P=\lambda/(hc)=1.91\times10^6$ noise photons per second at $\lambda=1$\,$\mu$m in the aperture of the SGL telescope. Compared to traditional observations using space telescopes, this is a large number: As shown in \citet{2017arXiv170603795H}, noise can be of order 0.1\,$\gamma$\,s$^{-1}$ for classical space telescopes.

\subsection{Methods to decrease noise}
Noise levels can be decreased by observing the Einstein ring further from the sun, at $45^{\circ}$ heliographic latitude one (two) orders of magnitude improvements are achieved for $b=1.35$, $z=1000$\,au ($b=2$, $z=2200$\,au).

Alternatively, larger telescopes (or shorter wavelengths) have better resolution, so that the ring-shaped coronal noise area decreases. As discussed in section~\ref{sec:coronograph}, a star shade can also help. Its working angle of 0.1\,arcsec is smaller than the diffraction limit of the single $d_{\rm SGL}=1$\,m telescope (0.25\,arcsec), and results in a smaller coronal noise area which overlays the Einstein ring. While the signal level stays the same, the reduction in noise is the same as for a 2.5\,m telescope ($A_{\circ}/A_{\odot}=0.14$ instead of 0.35). As will be discussed in section~\ref{sec_capacity}, this increases data rates by a factor of a few. Overall, the addition of a starshade is useful, but it is doubtful whether the added complexity and cost are worth it. In terms of noise, the combination of a 1\,m telescope and a 10\,m starshade is equivalent to a 2.5\,m telescope with a coronograph, and inferior to a 5\,m telescope with a coronograph.

Another method to decrease noise levels is to use narrower spectral filters. Laser line widths in the mHz range, although at low ($10^{-12}$\,W) power, have been demonstrated \citep{2009PhRvL.102p3601M,2012NaPho...6..687K}. Commercial high-power lasers at $\lambda=500$\,nm have line widths of 350\,MHz $=3\times10^{-4}$\,nm \citep{1999ApOpt..38.6347D}. Our telescope with a narrow 350\,MHz filter at $z=2200$\,au would collect a noise flux of 18\,$\gamma$\,s$^{-1}$, a comparably much lower number. This shows the strong positive influence on noise from heliocentric distance (through the coronal brightness power law scaling from the solar limb) and narrow bandwidth.

Additional encoding modes of photons are polarization and time (phase). Present (earth 2017) technology offers nanosecond observational cadence ($10^{-9}$~s) \citep{2007AcAau..61...78H}. We discuss the impact of these modes on data rate in section~\ref{sec_capacity}.

\section{Data rate}
\subsection{Capacity}
\label{sec_capacity}
The data rate of a noisy channel is the number of photons, multiplied with the capacity \citep[in bits per photon, as discussed in][]{2017arXiv170603795H}:
\begin{equation}
\label{bits_holevo}
\text{DSR}\textsubscript{$\gamma$} = C_{\rm th} F_{\rm r}~.
\end{equation}

The photon-limited capacity $C_{\rm th}$ of a noisy channel is \citep{2014NaPho...8..796G}
\begin{equation}
\label{thermal_holevo}
C_{\rm th}=g(\eta M + (1-\eta) N_{M}) - g((1-\eta)N_{M})
\end{equation}

in units of bits per photon, where $M$ is the number of photons per mode, $N_{M}$ is the average number of noise photons per mode, $\eta$ is the receiver efficiency and $g(x)=(1+x) \log_2 (1+x)-x \log_2 x$.

The capacity $C_{\rm th}$ is a logarithmic function of SNR. For example, if we have $\eta=0.5$ and $M=10^{-5}$ photons per mode, the maximum allowed noise is $N_M=0.13$ noise photons per mode in order to achieve an ordinary capacity of 1 bit per photon. A detailed discussion of the parameter space is in \citet{2017arXiv170603795H}.

\begin{figure}
\includegraphics[width=\linewidth]{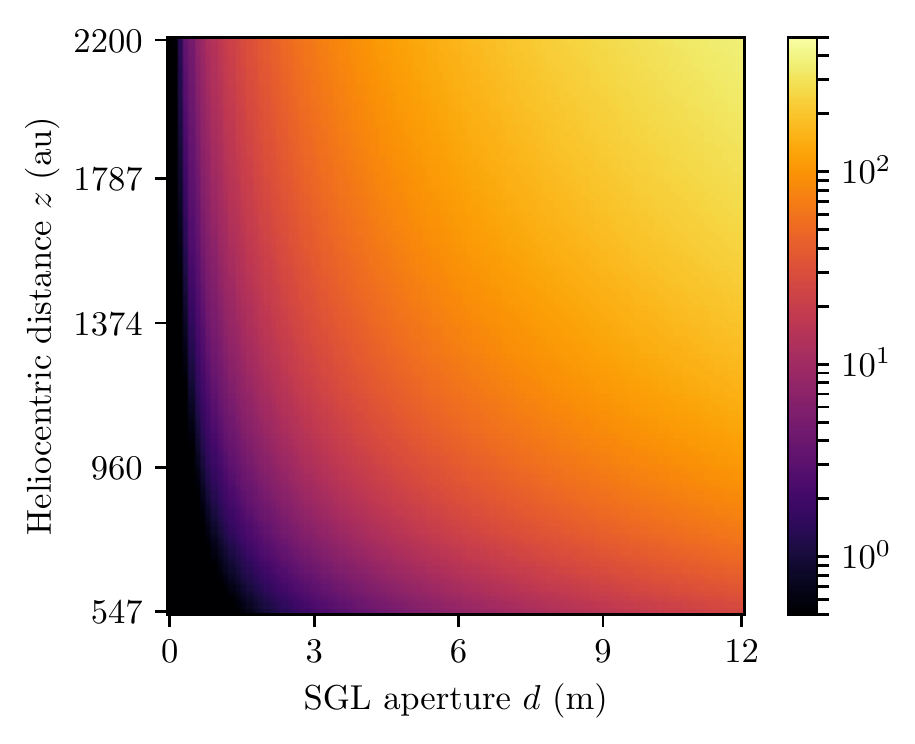}
\caption{\label{sgl_grid}Contours of data rate (in Mbits/s) for heliocentric distance $z$ and SGL telescope aperture $d_{\rm SGL}$ at $\lambda=1$\,$\mu$m. Increasing the aperture from $d_{\rm SGL}=1$\,m to 10\,m increases data rate $10\times$; moving the $d_{\rm SGL}=1$\,m telescope from $z=600$ to $z=2200$ increases data rate $2\times$.}
\end{figure}

\subsection{Data rate}
\label{sgl_aperture}
A major difference of the SGL telescope, compared to classical telescopes, is the much higher number of photons (both signal and noise) it receives. For the same capacity (in bits per photon), the number of modes to which these photons can be distributed must also increase.

As an example, consider our baseline scenario with a $d_{\rm SGL}=1$\,m telescope at $z=600$ observing a 1\,m probe at $\alpha$\,Cen ($R=1.3\,$pc) transmitting at 1\,W. With a technology offering $m=10^5$ modes, the capacity within the (physically allowed) Holevo bound is only $C_{\rm th}=0.027$ bits per photon. This low number comes from the comparably high number of (signal and noise) photons per mode ($M$), e.g. $S=3\times10^6\,\gamma\,$s$^{-1}$ so that $M=S/m\approx15$ for a data rate of 82\,kbits/s. The number of modes is the restricting factor in this regime where photons are plentiful. Increasing the number of modes to $m=10^{10}$ pushes the capacity to $C_{\rm th}=3.2$ bits per photon, for a data rate of 9.5\,Mbits/s. In this regime of high $m$ (order $10^{5}$), low $M$ (order $10^{-5}$ photons per mode), further increases in the number of modes are a logarithmic function of data rate, and only of marginal interest. An increase in transmitter power follows the inverse square law and requires only a low improvement in the number of modes, for a constant capacity. Very low transmitter power (so that $S \ll N$) causes an exponential penalty in capacity (and thus data rate) and should be avoided for efficient communication. In our example, this regime begins at $P<0.1\,$W.

The modes are discussed in greater detail in \citet{2017arXiv170603795H}. They originate from the photons' dimensions, namely polarization, frequency and time of arrival. For example, a number of $m=10^{10}$ modes can be realized with a combination of a $R_{\rm S}=100,000$ spectrograph and $10^5$ time slots. We use $m=10^{10}$ for the following example. Practical data rates using current technology are a few times below the quoted limits.

The strongest influence on data rate comes from $z$ and $d_{\rm SGL}$, and both are nonlinear. While the probe can achieve a data rate of 4.9\,Mbits/s at $z=600\,$au, the rate increases to 20.6\,Mbits/s at $z=2200\,$au, roughly a factor of two. An increase in the probe's aperture (1\,m to 10\,m) increases the data rate to 89\,Mbits/sec (at $z=600$\,au), roughly a factor of ten. Figure~\ref{sgl_grid} shows a grid for the ${\rm DSR_{\gamma}}(z, d_{\rm SGL})$ parameter space.

\subsection{Influence of wavelength}
\label{influence_of_wavelength}
For a constant noise level, data rate scales linear with frequency, so that decreasing wavelength from 1\,$\mu$m to 0.3\,$\mu$m (blue visible) increases data rate by $\approx3\times$, as noise is approximately constant in this band in the corona. Shorter wavelengths than 0.3\,$\mu$m are detrimental because of technological limitations \citep[][section 3.3]{2017arXiv170603795H}. Longer wavelengths are penalized by worse focusing, but profit from lower noise. For example, the number of photons received (signal) is $\approx3\times$ lower at 3\,$\mu$m compared to 1\,$\mu$m, but noise $\approx10\times$ lower, although the exact coronal flux is not known. We can estimate the capacity increase based on better SNR as $\approx20$\%, so that the loss of photons is not compensated, and data rates decrease with longer wavelengths.

An additional issue for infrared observations is the minimum telescope size to resolve the distance between the Einstein ring and the solar limb (section~\ref{sec:coronograph}). If the resolution is insufficient to resolve the gap, the full solar disk enters as noise, resulting in low data rates even for the relatively lower solar flux in these wavelengths. Moving the wavelength to THz and microwaves (the limit is at 120 GHz, section~\ref{sgl_frequency_cutoff}) further decreases noise, but the overall effect on data is also negative, driven by two factors: less efficient focusing for a constant aperture size, and the logarithmic nature of capacity as a function of SNR.

In other words, data rate does not benefit from the decrease in coronal noise at longer wavelengths, as long as power levels on the transmitter side are sufficiently high (of order $0.1 S >N$).

The ideal wavelength for the probe in the SGL is therefore at the technological limit: in the optical blue spectrum near 0.3\,$\mu$m. Interstellar extinction becomes relevant for distances $>200$\,pc \citep[][section 5.6]{2017arXiv170603795H} and increases the optimum wavelength with distance up to 3\,$\mu$m at 8\,kpc (the galactic center).

\begin{figure*}
\includegraphics[width=\linewidth]{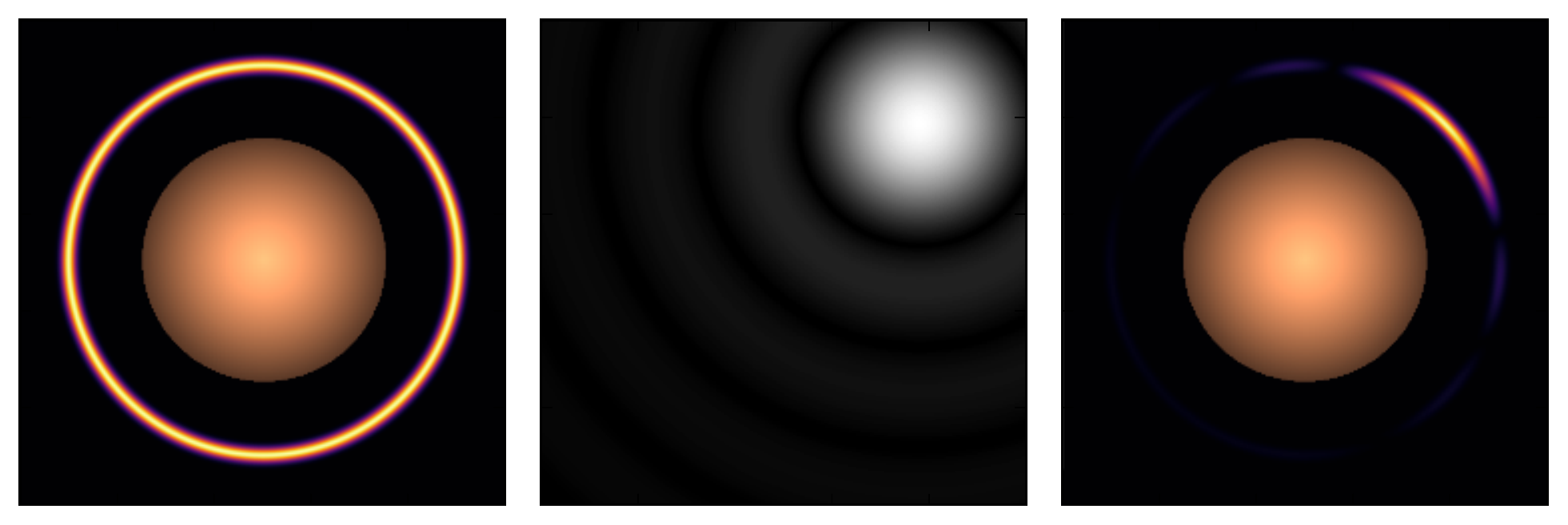}
\caption{\label{figure_gauss_lens}Left: View of the sun and evenly illuminated (on axis) Einstein ring for a receiving probe at $z=1400\,$au ($b=1.6\,R_{\odot}$). The brightness and width of the Einstein ring has been increased for better visibility. Middle: Airy pattern arriving ``behind'' the star (as seen from the receiver) with a beam width of $R_{\odot}$ to the first null, centered on the Einstein ring. Right: Einstein arc as a result of uneven illumination. Most of the flux (93\%) is centered on $\approx1/5$ of the circumference. The first airy ring is faintly visible in this color scale (blue: 2\% intensity).}
\end{figure*}

\subsection{Equivalent technology}
We will now examine technological equivalents to achieve the same data rate without an SGL probe. This is useful to determine the benefits of an expensive probe, and possible replacements with standard equipment.

A crucial parameter in these estimates is the number of modes $m$. This parameter is determined by technology, and a detailed analysis of earth 2017 capabilities is beyond the scope of this paper. We will discuss the optimistic, and the pessimistic case. A number of $10^5$ modes should be comparably easy to implement in a single spectral channel with clock synchronization better than $10^{-5}$. A higher number of $10^{10}$ modes might require the addition of a spectrograph, and be comparably ambitious.

For $m=10^5$, the standard SGL probe receives a data rate of 82\,kbits/s per Watt. To achieve the same data rate with a large classical telescope (E-ELT with $D_{\rm classical}=39\,$m), the power level of the transmitter needs to increase from 1\,W to 100\,kW. If the power level remains at 1\,W, the classical aperture needs to have a diameter of 13\,km for the same data rate (the photon-collection equivalent is 75.75\,km, but the SGL telescope collects much more noise).

For $m=10^{10}$, the standard SGL probe receives a data rate of 9.5\,Mbits/s per Watt. This would require transmitter power levels of MW ($10^6$\,W) when matched with an E-ELT, or a classical telescope of 71\,km size for a 1\,W communication. With $m=10^{10}$, the SGL telescope realizes a significant fraction (88\%) of its data rate capabilities (calculated from its aperture), despite coronal noise.

Increasing $d_{\rm SGL}$ or moving the telescope further out in $z$ produces data rates which are only obtainable with classical telescope using power levels or megastructures that are unrealistically large for earth 2017 technology.

\section{Einstein rings versus arcs}
\label{einstein_arc}
The Einstein ring configuration which is typically considered has the light source as an isotropic emitter at a large distance from the source. The source illuminates the gravitational lens evenly, although there might be a misalignment of source and lens along our line of sight. 

In the case of our probe, the source is neither infinitely far away nor radiating isotropically. Instead, we have assumed the probe at a distance of 1.3\,pc transmitting into a tight cone limited by diffraction. At this distance, the beam width of a $D_{\rm t}=1\,$m telescope transmitting at $\lambda=1\mu$m is $\approx 2\,{\rm au} \approx 435\,R_{\odot} \gg R_{\odot}$. In this regime, the approximation holds and no special treatment is required.

For larger transmitters, shorter distances and/or shorter wavelengths, the beam width decreases. As an example, in the same configuration a wavelength of $\lambda=2$\,nm (or a transmitter with $D_{\rm t}=435\,$m) produces a beam width of $\approx 1\,R_{\odot}$. If the beam is centered on the star, the first null of the Airy pattern overlays the Einstein ring. Then, zero flux is injected into the gravitational lens.

We have calculated the optimal placement of the beam center as a function of beam width. As shown in Figure~\ref{figure_gauss_lens}, we vary the center of the Airy pattern and its width and count the photons collected by the gravitational lens by numerically integrating over the thin strip which forms the Einstein ring. If the center of the beam is on the ring, and the beam is not much larger than the star, an Einstein arc is formed (Figure~\ref{figure_gauss_lens}, right).

\begin{figure}
\includegraphics[width=\linewidth]{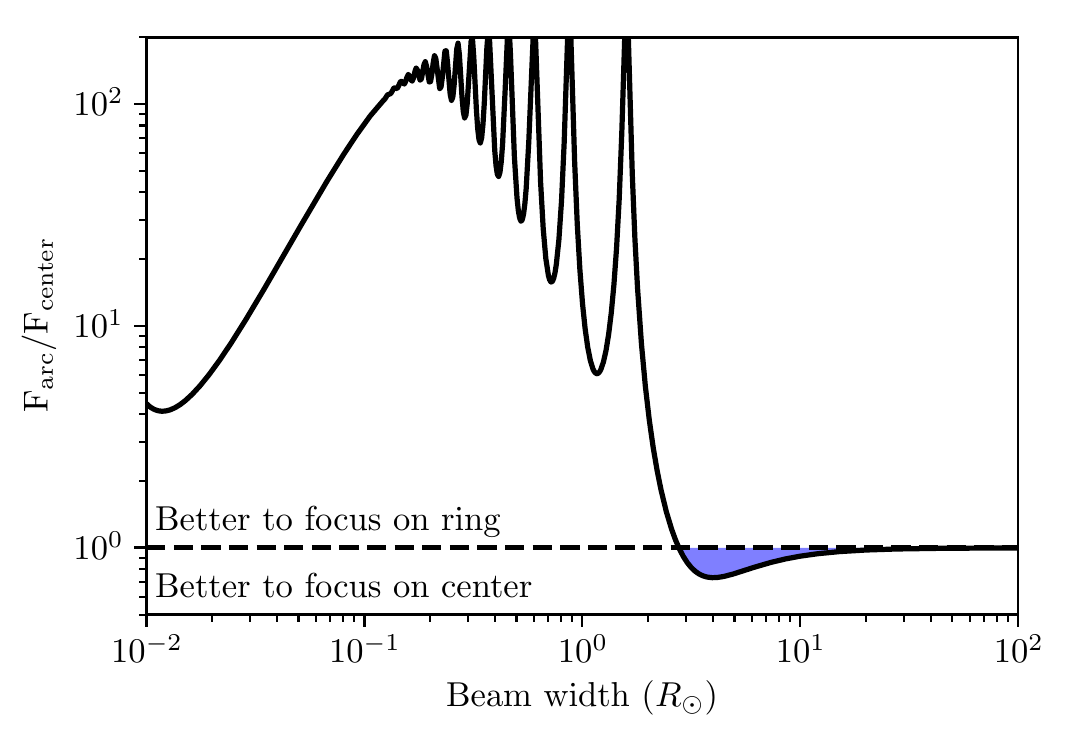}
\caption{\label{figure_lens_focus}Ratio of flux retrieval of a beam focused on the ring and a beam focused on the center of a star. For beam widths $\lesssim
3R_{\odot}$ (at $b=1.4$), it is beneficial to aim the beam at the ring; larger beams should be centered on the star. The difference is $<10\%$ for beam widths $\gtrsim 9R_{\odot}$.}
\end{figure}

Our numerical simulations show that it is optimal to place the beam center on the ring for beam widths $\lesssim 3R_{\odot}$ (at $b=1.4$ used in this example). Larger beams should always be centered on the star. The difference between both placements becomes $<10\%$ for beam widths $\gtrsim 9R_{\odot}$ (Figure~\ref{figure_lens_focus}).

\section{Discussion}
\subsection{Previous mission designs}
Several SGL mission concepts have been presented in the literature. The FOCAL mission is a concept proposed to ESA in 1993 \citep{1994JBIS...47....3M} and refined in a series of more than 20 papers. It suggests an interstellar radio link using frequencies between 327\,MHz \citep{2010AcAau..67..521M,2010AcAau..67..526M,2013AcAau..82..246M}, 1.4 and 22\,GHz with a 12\,m telescope \citep[e.g.][]{2010AcAau..67..521M}.

The idea was expanded to using the telescope for observations of the cosmic microwave background near its peak at 160.378\,GHz \citep{2000AcAau..46..605M}, or by using a ``double bridge'' of two 10\,m telescopes in the SGLs of two stars \citep{2011AcAau..68...76M}.

In a NASA-sponsored study, \citet{1999AcAau..44...99W} discusses nuclear propulsion and solar sails, and shows SGL telescope designs ranging from 1\,m to 24\,m, covering frequencies (wavelengths) from microwaves (3 GHz) to ultraviolet (300\,nm).

All of these studies should be regarded as design ideas, rather than serious mission concepts. First, frequencies $<122$\,GHz have no focus in the SGL due to coronal electron deflection (section~\ref{sgl_frequency_cutoff}), rendering some of the ideas completely unfeasible.

Second, the gain in the image field is highly non-uniform (section~\ref{sub_light_collection_power}), which makes all previous gain calculation over-estimates by orders of magnitude. Third, as a result from incorrect gain calculations, the proposed telescope sizes (e.g. 10\,m) are arbitrary.

Fourth, it has been suggested to use a communication scheme of two 10\,m sized probes in the gravitational lens of two stars to multiply the gain \citep{2011AcAau..68...76M}. This is impossible because for telescope diameters $<14.8$\,km (at $\lambda=1$\,$\mu$m), the Einstein ring is unresolved, and the majority ($4.6\times10^{-9}$, section~\ref{sub_lens_geometry}) of the flux would not be transmitted into the lens. Transmitting on the direct path would have the same outcome, and avoid the difficulties of having the sun in the focus.

\subsection{Required precision for the placement of the probe}
A probe in the SGL needs to be placed, and kept in place, in the focus very accurately. The focus is a caustic line, so that a movement in heliocentric distance $z$ is acceptable, even beneficial. The required accuracy on the axis was already estimated for the PSF width (6\,dB) as 14\,m at 300\,GHz, by \citet{2011JBIS...64...24G}. This scales linearly to 14\,cm at $1\,\mu$m and is in principle agreement with the correct PSF width (to the first null, section~\ref{sgl_aperture}) of 4.7\,cm (at $1\,\mu$m for $z=600$\,au), so that the spacecraft needs to be positioned, and permanently guided, within cm to m from the focus. The side-lobes which decrease in power with distance from the axis (Figure~\ref{figure_bessel}) can help to find the focus.

\subsection{Pointing accuracy of the probe}
The distant probe needs to point its beam at the sun or the Einstein ring (section~\ref{einstein_arc}). For the large beam width of $\theta\approx 0.25$\,arcsec in our standard example, the Airy disk can be well approximated with a Gaussian. If we want to limit the acceptable flux loss from pointing errors to e.g. 10\% results at the receiver, a pointing accuracy of the same order (25\,mas) is required. This is trivially possible with a meter-class space telescope. For comparison, the Gaia mission will achieve astrometry at the $\mu$as level \citep{2012Ap&SS.341...31D}.

\subsection{Orbital stability}
\label{orbital_stability}
A probe moving away in $z$ improves its data rate as long as it can stay exactly on axis. Communication back to earth, however, gets more difficult with increasing distance. If the plan is to keep the probe at a constant distance $z$ on the lens axis, it needs to compensate the gravitational pull of the sun (we neglect here the question of how the probe decelerates). Without an internal energy source, a solar sail could be used. Assuming an ideal flat sail with 100\% reflection, the critical sail surface density (including payload) to balance photon pressure with gravity is 1.4\,g\,m$^{-2}$ \citep{2014AcAau..94..629G}. At 600\,au, the solar pressure is 3.8~mW\,m$^{-2}$, causing a force on the meter-sized sail of $P=2.5 \times 10^{-11}$\,N. With an additional force of the same magnitude, perhaps through harvesting solar power with a very lightweight collector, an acceleration of $\approx0.5$m\,s$^{-1}$\,yr$^{-1}$ can be achieved. This way, the probe could travel $\leq 10,000$\,km per year, $\leq 10^{-7}$ of a revolution around the Sun at 600\,au. With such low power, course corrections to compensate for the relative position of the star at the other end of the lens are possible except for stars with proper motions $>10$~arcsec yr$^{-1}$, but moving the probe to target other stars would take very long. For comparison, the proper motion of $\alpha$\,Cen~AB is 3.7\,arcsec\,yr$^{-1}$ \citep{2016A&A...594A.107K}. Thus, a probe that is once stationed at constant $z$ would likely forever target the same object. This assumes that the receiving probe does not rotate around the sun in the first place. If the source (or transmitter) is in orbit around another star (either for communication of imaging of exoplanets), this motion needs to be compensated for. An orbit of 1\,au at a distance of 1.3\,pc (Proxima Cen) translates to an ellipse with semimajor axis of $150,000$\,km in the image plane.

\begin{figure*}
\includegraphics[width=.5\linewidth]{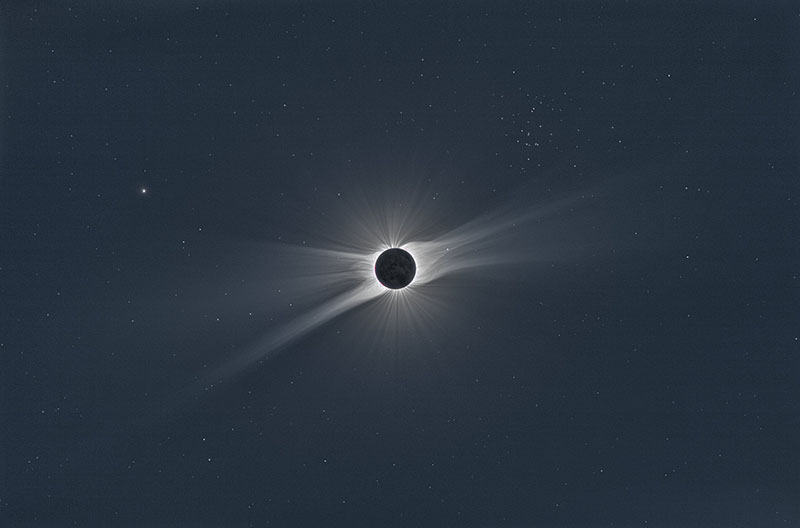}
\includegraphics[width=.5\linewidth]{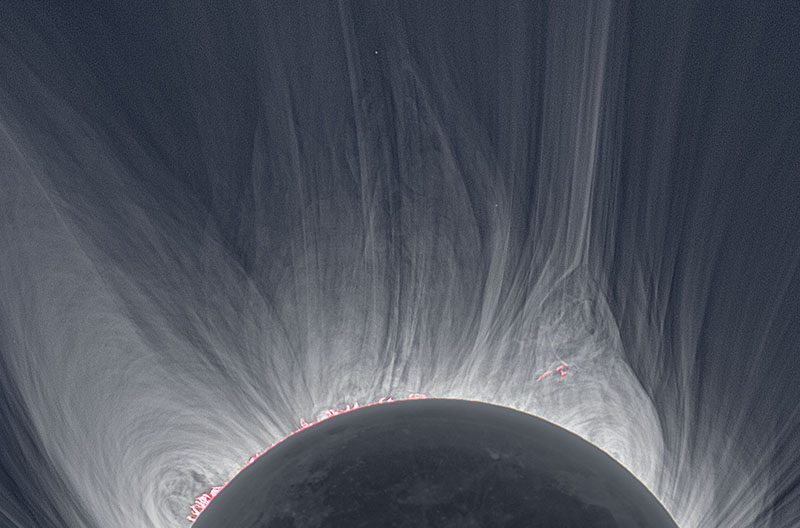}
\caption{\label{fig_corona}Structure in the solar corona in the optical. Images reproduced with permission from Miloslav Druckm\"uller.}
\end{figure*}

\subsection{Signal-to-noise ratio of exoplanet imaging}
The high resolution of the SGL telescope (section~\ref{sub_resolution}) is beneficial for exoplanet imaging as it places the host star \citep[with $>10^9$ higher flux,][]{2010exop.book..111T} outside of the field of view when observing the exoplanet. As an example, the flux (on the direct path) from our nearest neighbor star Proxima Centauri ($L=1.38\times10^{-4} L_{\odot}$) is $4.25\times10^{6} \gamma$\,sec$^{-1}$\,m$^{-2}$. For an optimistic exoplanet contrast of $10^{-9}$ it is $4.25\times10^{-3} \gamma$\,sec$^{-1}$\,m$^{-2}$. The major problem here is that the signal is not in a narrow band, but spread over the whole spectrum, just as the coronal noise. Exoplanet flux peaks in the micron wavelengths \citep{2010exop.book..111T} just as the corona. Good SNR ($>1$) is only achievable for the most favorable SGL telescope configurations. For example, at $z=600$\,au, $d_{\rm SGL}=1$\,m, SNR$=2\times10^{-3}$, which improves with $z$ and $d_{\rm SGL}$ (as shown in Figure~\ref{sgl_grid}) to SNR$=38.2$ for $z=2200$\,au and $d_{\rm SGL}=10$\,m. The broadband flux from Proxima~b has the same order-of-magnitude SNR in an SGL telescope as a 1\,W laser from this planet in a 1\,nm channel.

\subsection{Aberrations from the oblateness of the sun}
The sun is not a perfect sphere; it is oblate because of rotation by $8\times10^{-6}R_{\odot}$ \citep{2012Sci...337.1611G}. The corresponding shift of the focal plane is of order 30\,cm towards the solar equator \citep{2005ExA....20..307K}, and the influence on the PSF itself is negligible. Similar gravitational shifts are to be expected from the orbits of the planets (of order meters), and the probe will need to compensate for the combined gravitational shifts as caused by all bodies in the solar system.

\subsection{Evolution of solar coronal noise}
As the probe observes through the solar corona, it picks up its noise photons (section~\ref{sub_corona}). This noise depends on the wavelength and the heliocentric latitude (Figure~\ref{figure_corona}), and its structure changes over time. Figure~\ref{fig_corona} shows the situation for one specific time in the optical. Encoding schemes will need to account for variations in noise, so that transmission is still possible at all times (transmission occurs 4.25 years before receiving, so that the actual noise levels at the time of arrival are unknown).

\subsection{Communication between the SGL probe and Earth}
If the probe acts as a relay (i.e. all communication it receives is simply forwarded to earth), and the data rates are DSR$_{\rm dist}$ and DSR$_{\rm earth}$, the duty cycle (fraction of total time available for interstellar communication) is

\begin{equation}
{\rm DC} = \frac{1 - 2{\rm t}_{\rm switch}-{\rm DSR}_{\rm dist}}{{\rm DSR}_{\rm earth}+{\rm DSR}_{\rm dist}}.
\end{equation}

Following our calculation framework in \citep{2017arXiv170603795H}, the data rate between a $d_{\rm SGL}=1$\,m telescope at $z=600$\,au and an E-ELT on earth ($D_{\rm classical}=39$\,m) is 320\,kbits per second per Watt, which is about one order of magnitude lower than the communication through the SGL with the same aperture size, wavelength and power. For a duty-cycle of 50\%, the probe in the SGL needs to increase its power (towards earth) by $100\times$ compared to the transmitter power at $\alpha$\,Cen. This shows the enormous gain of the SGL.

\subsection{ET probes}
It has been suggested to search for ET probes in the SGL \citep{2014AcAau..94..629G}. As there are an infinite number of foci in a sphere around our sun, the problem is to select the communication target. A priority list might include the nearest stars, nearby stars with habitable exoplanets, the nearest exoplanets with biological life (of which none are known to us so far), and the galactic center. A detailed discussion will be given in paper III of this series.

\subsection{\texttt{PyCom} software package}
We provide the Python-based software package \texttt{PyCom} as open source under the free MIT license\footnote{\hyperref[]{http://github.com/hippke/communication/}, commit {\tt 592a28d}}. It offers function calls for the equations in this paper, and example files to reproduce the calculations and figures.

\section{Conclusion and outlook}
We have shown, for the first time, that the gravitational lens of our sun can be effectively used for interstellar communication. This had previously been unclear due to the unknown impact of the coronal noise. We have calculated point-spread functions, aperture sizes, heliocentric distance, and optimum communication frequency of a receiving probe in the SGL.

The requirement to accurately position the spacecraft in the image plane is difficult, because it requires a precision of $\approx10$\,cm in lateral directions at $\lambda=1\,\mu$m. If an occulter is used, this second spacecraft would need to be in formation flight 78\,km from the receiver, and $\approx10$\,m in size.

Data rates are higher by a factor of $10^7$ compared to equal-sized classical telescopes. A 1\,m telescope in the SGL can achieve the same receiving data rate as a classical 13--75\,km telescope. If classical telescope sizes are restricted to E-ELT size (39\,m), power levels on the transmitter side need to increase from 1\,W into the MW range to match the data rate of an SGL probe.

If data rates at Mbits/s level are required, it might be cheaper to invest into space flight to the SGL. The obvious alternative is to limit data rates to a level achievable with smaller telescopes. A single 39\,m telescopes may receive data of order bits per second per Watt from $\alpha$\,Cen, sufficient to transmit several high resolution photographs over the course of a year \citep{2017arXiv170603795H}. Perhaps this is judged to be sufficient.

In paper III of this series, we will relax technological constraints, mainly on the focusing of short wavelengths \citep{2017arXiv170603795H}. This opens our horizon to more advanced civilizations, if they exist, and allows us to examine how they would maximize data rates. If advanced civilizations value data as much as we do, our framework will tell us how they communicate, where we can look for such communication, and how we could join the galactic network.

\acknowledgments
\texttt{Acknowledgments.} The author is thankful to Slava Turyshev and Duncan Forgan for helpful discussions, and to the Breakthrough Initiatives for an invitation to the Breakthrough Discuss 2017 conference at Stanford University.

\bibliographystyle{yahapj}

\end{document}